LA-UR-12-23257

Title: Current status of MCNP6 as a simulation tool useful for space and accelerator applications

Author(s): Mashnik, Stepan G
Bull, Jeffrey S
Hughes, H. Grady
Prael, Richard E
Sierk, Arnold J

Intended for: 11th Conference on the Intersections of Particle and Nuclear Physics (CIPANP 2012), 2012-05-28/2012-06-03 (St. Petersburg, Florida, United States)

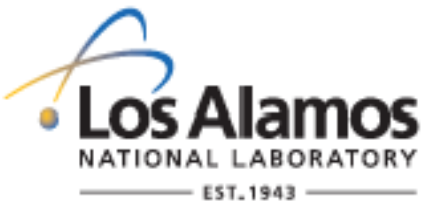

# Current status of MCNP6 as a simulation tool useful for space and accelerator applications

S. G. Mashnik, J. S. Bull, H. G. Hughes, R. E. Prael, A. J. Sierk

*Los Alamos National Laboratory, Los Alamos, NM 87545, USA*

**Abstract.** For the past several years, a major effort has been undertaken at Los Alamos National Laboratory (LANL) to develop the transport code MCNP6, the latest LANL Monte-Carlo transport code representing a merger and improvement of MCNP5 and MCNPX. We emphasize a description of the latest developments of MCNP6 at higher energies to improve its reliability in calculating rare-isotope production, high-energy cumulative particle production, and a gamut of reactions important for space-radiation shielding, cosmic-ray propagation, and accelerator applications. We present several examples of validation and verification of MCNP6 compared to a wide variety of intermediate- and high-energy experimental data on reactions induced by photons, mesons, nucleons, and nuclei at energies from tens of MeV to about 1 TeV/nucleon, and compare to results from other modern simulation tools.

**Keywords:** MCNP6, Monte-Carlo method, spallation, fragmentation, evaporation, fission, CEM03.03, LAQGSM03.03, INCL+ABLA, ISABEL, Bertini+MPM+Dresner+RAL
**PACS:** 72.10.Fk, 29.20.-c, 61.82.Bx, 24.10.Lx, 25.40Sc, 25.70, 25.85.-w

## INTRODUCTION

The Los Alamos National Laboratory (LANL) particle radiation transport code MCNP, which stands for Monte-Carlo N-Particle, is a general purpose three dimensional simulation tool that transports 37 different particle types and arbitrary heavy ions for different applications, including space and accelerator.

Monte-Carlo particle radiation transport methods have had an extensive history at LANL dating from the 1940s. Early creators of these methods include Drs. Stanislaw Ulam, John von Neumann, Robert Richtmyer, Nicholas Metropolis, and others, who investigated neutron transport issues on first-generation computers. On March 11, 1947, John von Neumann sent a letter to Robert Richtmyer, leader of the Theoretical Division at Los Alamos, proposing the use of the statistical method to solve neutron diffusion and multiplication problems in fission devices. His letter was the first formulation of a Monte Carlo computation for an electronic computing machine (see references and more details in Ref. [1]). In 1947, while at Los Alamos, Fermi invented a mechanical device called FERMIAC11 to trace neutron movements through fissionable materials by the Monte-Carlo Method. During the 1950s–1960s, a number of special-purpose Monte-Carlo codes were developed at LANL, including MCS, MCN, MCP, and MCG. These methods eventually found their way into a code called MCNG, which was first created in 1973 by merging a three dimensional neutron-transport code MCN, with the $\gamma$-transport code MCG. In 1977 MCNG was merged with MCP, a Monte-Carlo photon code with detailed physics treatment down to 1 keV, to more accurately model neutron-photon interactions. The resulting code, MCNP, originally stood for Monte Carlo Neutron Photon. In 1983, MCNP3 was released for public distribution to the Radiation Safety Information Computational Center (RSICC) at Oak Ridge, USA.

The meaning of MCNP changed to Monte-Carlo N-Particle when electron transport, from Sandia National Laboratory's Integrated TIGER Series (ITS) was added in 1990. MCNP has been expanded ever since to include more and more particle types. In 1996, the LANL code LAHET was added to MCNP4B, creating a "Many-Particle MCNP Patch". The utility of many-particle transport has found many applications and sponsors, and continued to grow as a separate code, MCNPX [2]. In 2001–2002 MCNP4C was completely rewritten in modern Fortran 90, and was enhanced to support large scale parallelism using combined MPI message passing and OpenMP threading, resulting in MCNP5 [3]. In July 2006, a merger effort was started, taking MCNPX 2.6.B and adding it to a LANL version of MCNP5. The resulting code, MCNP6, took more than twelve man-years to create from the two parent codes. Fig. 1 shows a simplified, schematic history of MCNP6.

## MCNP6 MODELS, ITS V&V AND USE

MCNP6 considers several nuclear-reaction models, sometimes incorporated in separate modules we refer to as "event-generators". The first model of nuclear reactions used initially in LAHET [4] was the Bertini INC [5] followed by the Multistage Preequilibrium Model (MPM) [6] followed by the evaporation model as implemented in the code EVAP by Dresner [7]. If the compound nuclei produced after the INC and MPM stages of reactions are heavy enough to fission, the fission process is simulated either with the semi-phenomenological Atchison fission model, often referred in the literature at the Rutherford Appleton Laboratory (RAL) fission model, which is where Atchison developed it [8], or with the Fong statistical model of fission as implemented in the ORNL code HETFIS [9], often referred in the literature as the ORNL fission model. Bertini INC, MPM, EVAP, RAL, and HETFIS migrated from LAHET to MCNPX and later, to MCNP6.

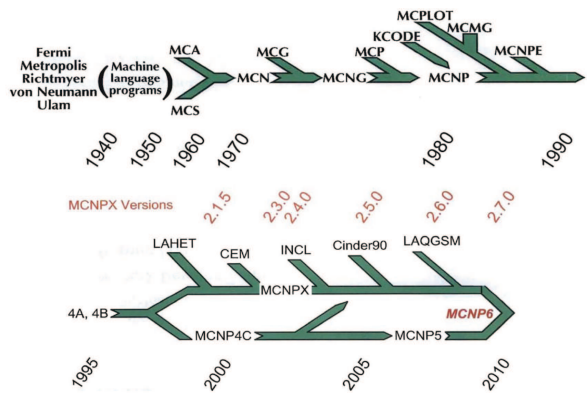

**FIGURE 1.** A schematic history of MCNP6.

The second model, from a historical point of view, which migrated to MCNP6 via MCNPX from LAHET is the ISABEL INC [10]. Just like the Bertini INC, ISABEL can be used with or without taking into account preequilibrium reactions as described by MPM and it can describe the evaporation and fission reactions with EVAP, RAL, and HETFIS.

A newer and recently improved model used by MCNP6 is the Cascade-Exciton Model (CEM) of nuclear reactions as implemented in the event-generator CEM03.03 [11, 12]. CEM03.03 uses its own models to describe the cascade, preequilibrium, evaporation, and fission reactions. It considers also coalescence of cascade nucleons into complex particles up to $^4$He and Fermi break-up of excited or unstable nuclei with mass numbers up to A = 12 (see details in Refs. [11, 12]).

Another new and recently improved model used by MCNP6 is the Intra-Nuclear-Cascade model developed at Liege (INCL) by Joseph Cugnon et al. [13]. INCL always uses only the ABLA code developed at GSI [14] to describe the evaporation and fission stages of reactions, independently of what MCNP6 users would chose for evaporation/fission models; INCL does not consider preequilibrium reactions. Newer and better versions of INCL and ABLA are planned to be incorporated into a future version of MCNP6.

Finally, MCNP6 uses the Los Alamos version of the Quark-Gluon String Model (LAQGSM) as implemented in the event-generator LAQGSM03.03 [11, 15]. LAQGSM was developed to describe reactions induced by almost all types of elementary particles and by nuclei at energies up to about 1 TeV/nucleon. LAQGSM uses its own models to describe the cascade, preequilibrium, evaporation, and fission reactions; it considers also coalescence of cascade nucleons into complex particles up to $^4$He and Fermi break-up of excited or unstable nuclei with mass numbers up to A = 12 (these are the same models used by CEM, but adjusted to LAQGSM; see details in Refs. [11, 15]).

Before distributing MCNP6 to the public, we have tested and validated it on as many test problems as possible, using reliable experimental data. Extensive Validation and Verification (V&V) has been performed and documented in many publications (see, e.g. [1, 16, 17] and references therein). Fig. 2 presents only one example from Ref. [16]. We obtained similar results for many other reactions of interest for FRIB/RIA and space applications (see more details in Ref. [16]).

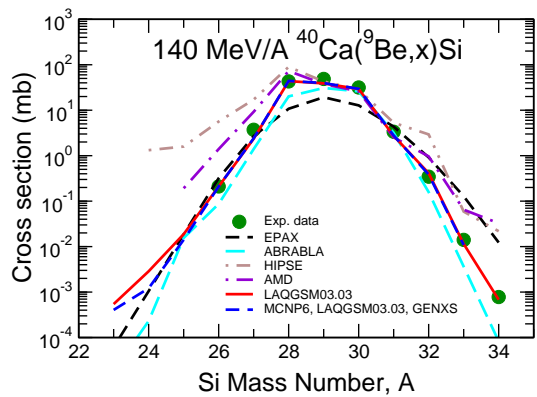

**FIGURE 2.** Experimental [18] mass-number yields of Si isotopes produced from 140 MeV/A $^{40}$Ca + $^9$Be (green filled circles) compared with results from EPAX [19], ABRABLA [14], HIPSE [20], and AMD [21] from [18], as well as with predictions by LAQGSM03.03 used as a stand alone code and by MCNP6 using the LAQGSM03.03 event-generator, as indicated.

Naturally, during our extensive V&V work of MCNP6, we discovered some "bugs" and more serious physics problems in MCNP6 or/and in MCNPX/5. Most of them have been fixed (see examples in Refs. [1, 16, 17]). We continue our work to solve all the observed problems before a "production" version of MCNP6 is distributed to the public. Fig. 3 presents only one recent example. A previously unobserved error in the calculation of fission cross sections of $^{181}$Ta and other nearby target nuclei by the CEM03.03 event gener-ator in MCNP6 and a technical "bug" in the calculation of fission cross sections with MCNP6 while using the LAQGSM03.03 event generator were detected and fixed in Ref. [17].

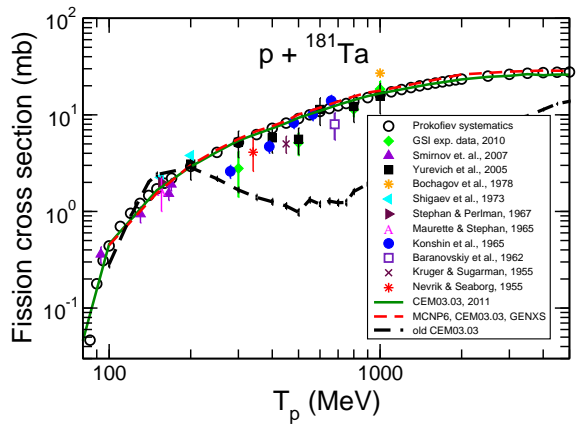

**FIGURE 3.** Prokofiev systematics [22] (open circles) and experimental proton-induced fission cross sections of $^{181}$Ta (symbols, see detailed references in [17]) compared with our old MCNP6 calculations (black dashed lines) using the CEM03.03 event generator before we fixed the error, with the corrected CEM03.03 results (green line), and with calculations by the updated MCNP6 using the corrected CEM03.03 event generator (red dashed line), as indicated.

After fixing these problems, we find that MCNP6 using the CEM03.03 and LAQGSM03.03 event generators calculates fission cross sections in a good agreement with available experimental data for reactions induced by nucleons, pions, and photons on both subactinide and actinide nuclei at incident energies from several tens of MeV to about 1 TeV.

MCNP6 and its precursors MCNPX/5 have been used for many years in many space and accelerator simulations. They proved to be convenient and very useful tools for such applications (see, e.g. [23]–[25] and refrences therein).

## CONCLUSIONS

MCNP6 has been validated and verified against a variety of intermediate and high-energy experimental data of interest to space and accelerator applications and against results by several other codes. We find that MCNP6 describes reasonably well various reactions induced by particles and heavy ions at incident energies from their thresholds up to $\sim 1$ TeV/nucleon, measured on both thin and thick targets, and agrees very well with similar results obtained with MCNPX and other codes. MCNP6 and its precursor MCNPX have been already used succesfuly in different space and accelerator simulations and proved to be useful tools for such applications. The "Beta 2" version of MCNP6 is available to the public via RSICC at Oak Ridge, USA. We plan to deliver a "production" version of MCNP6 to RSICC during FY2013.